\documentclass[a4paper]{jpconf}
\usepackage{graphicx}
\usepackage{amsmath,amssymb,amsfonts}
\usepackage{booktabs}
\usepackage{array}
\usepackage{enumerate}
\usepackage{cite}
\usepackage{hyperref}

\def\beq{\begin{equation}}
\def\be{\begin{equation}}
\def\ee{\end{equation}}
\def\bes{\begin{eqnarray}}
\def\ees{\end{eqnarray}}

\def\f{\frac}

\def\pp{\partial}

\def\R{{\mathcal R}}

\def\calO{{\mathcal O}}

\begin{document}
\title{Excavations at the gravitationally collapsed site: Recent findings\footnote{Based on the talk `Black holes: Up, close and personal', I gave at ERE 2014, Valencia, Spain.}}

\author{Suprit Singh}

\address{Inter-University Centre for Astronomy and Astrophysics, Ganeshkhind, Pune 411 007 India}

\ead{suprit@iucaa.ernet.in}

\begin{abstract}
Hawking effect was dug out of the gravitationally collapsed site forty years back when it was realised that quantum effects at the horizon could propagate outward to infinity giving rise to a thermal flux at late-times. However, the situation regarding non-asymptotic observers was never completely clear. Also, recently a debate has sprung in the community as to what would infalling observers perceive while crossing the horizon. We set out to settle this question and more at least semi-classically in the articles [arXiv:1304.2858] and [arXiv:1404.0684] with a fresh approach. We introduce a new set of coordinates that are regular everywhere, consider the adiabatic expansion of detector response and its link to the trajectory-dependent `effective' temperature/s and also local invariant observables, energy density and flux, built from renormalized stress energy tensor. This paper is a concise summary of the new procedure and the results obtained thereof.
\end{abstract}

\section{Introduction}

\begin{figure}[t!]
\centering
\label{penrosecol}
\includegraphics[scale=0.81]{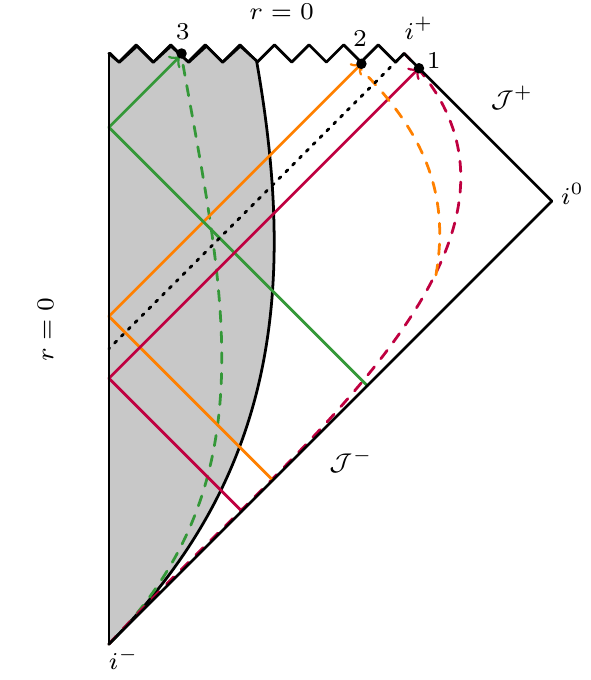}\hspace{15pt}
\includegraphics[scale=0.79]{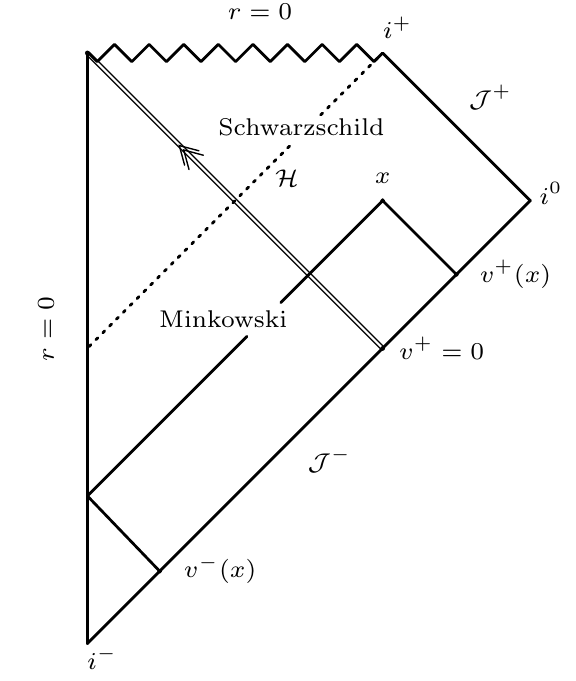}
\caption{{\bf Left.} Penrose diagram for a generic gravitational collapse scenario showing three different null rays and observers. The static observers outside correspond to reaching point $1$, point $2$ corresponds to the radial infallers while those inside the collapsing body reach point $3$ presenting different aspects of the Hawking phenomenon. {\bf Right.} Construction of globally defined null coordinates. Any point $x$ on the spacetime can be ascribed two coordinates $(v^+,v^-)$ which correspond to one ray coming directly from $\mathcal{J}^-$ and the other traced back to $\mathcal{J}^-$ after a reflection from the origin. Here, we take Vaidya null collapse where the thin null shell lies at $v^+ = 0$ separating the Minkowski interior from the Schwarzchild exterior.}
\end{figure}

Hawking famously showed \cite{Hawking74} that black holes emit a stationary thermal flux with temperature $T_H = 1/8\pi M$  which is perceived by the asymptotic observers at late-times. But that is not all that happens in the system that undergoes a gravitational collapse. For example, in a typical collapse scenario (see Fig.~\ref{penrosecol} (Left)), the usual Hawking effect is understood due to the `peeling' of modes that travel close to the horizon just before the collapse and escape to infinity with an infinite redshift. This is depicted by ray number $1$ and the corresponding static observer encounters a temperature with the standard Tolman shift which becomes unity in the asymptotic limit. However, we can have a similar null ray numbered $2$ here, which would be just inside the horizon and hits the singularity at some finite time. The infallers (radial or otherwise) would encounter this situation. The third possibility is of the ray $3$ that is inside the collapsing star all the time and here we would have comoving observers (the inside spacetime being Friedmann for dust collapse). Thus, we can have many different observers/frames which \emph{should} see the effect of collapse on the quantum vacuum differently and there is a rich structure beneath waiting to be explored.   

On a related note, a class of observers termed the \emph{infallers} have recently sprung a hot debate in the community on what would happen \emph{at the horizon crossing} \`a la the Hawking effect. It is usually argued \cite{Unruh76,Wilburn11} on the basis of equivalence principle that the infalling observer should see a quantum vacuum state while crossing the horizon and nothing special. However, this line of thought has been questioned lately by AMPS \cite{Almheiri13} claiming on the quantum-information-theoretic grounds that the infalling observers will encounter a \emph{firewall} at the horizon. Amidst these speculations which are not completely understood at this time, we seek to know what happens at least \emph{semi-classically}, relying on the well-established grounds. 

Working on these lines, Barbado et al \cite{Barbado11} introduced the notion of an `effective' temperature as a measure of the response of an Unruh-Dewitt detector \cite{Unruh76, DeWitt79} moving on a given trajectory. They found that the effective temperature increases for a detector dropped from infinity on a radially infalling geodesic as the infall progresses and reaches the temperature \emph{four} times the Hawking temperature at the horizon. As much as this result is intriguing, it should be noted their analysis does not give any direct link with the detector response nor does it provide any understanding to how thermal or non-thermal the effect is. Another point of concern is the choice of vacuum which they choose to be some new unconventional state in extended Schwarzchild geometry. We might think the peculiar result could be an artefact of this contraption. But more importantly, the use of both the ingoing and outgoing Eddington-Finkelstein coordinates together is not suitable if we wish to study the interior of the black hole horizon.

With this retrospection of the cards on the table, we now proceed to bring in a new deck \cite{ms2013,ss2014a} with the following assumptions and characteristics: (a) We introduce a new set of globally defined coordinates which are regular everywhere. (b) Working with the massless scalar field, the vacuum state in a collapse scenario is a natural and uniquely defined in-vacuum on $\mathcal{J}^-$. (c) We define and study a suitably defined time-dependent `temperature' function also relating it explicitly to UDW response functions by an \emph{adiabatic expansion}
and finally (d) We look at the observables -- energy density and fluxes -- built from the renormalized stress energy tensor (RSET) which explicitly depend on observer's kinematics for an alternate description. This also confirms the results of the effective temperature approach. 

This paper is meant to give a concise description of the formalism and results obtained in the previous works \cite{ms2013,ss2014a} which relate with the collapse of a thin null shell (ingoing Vaidya geometry) and to provide a preview of the results (relating to time-like dust collapse) which are a work in progress \cite{ss2014b}.
  
\section{The new procedure}

Without further ado, we shall present the new procedure and its application to the thin null shell collapse. As remarked earlier, it begins with the introduction of the \emph{new} coordinates.
\begin{itemize}
\item[(1)] {\bf Coordinatization:} Any point $x$ on the spacetime manifold in the collapse scenario (Fig.~\ref{penrosecol} (Right)) can be ascribed two coordinates $(v^+,v^-)$ which correspond to one ray coming directly from $\mathcal{J}^-$ and the other traced back to $\mathcal{J}^-$ after a reflection from the origin. This construction is completely general and valid for any type of collapse. For the Vaidya collapse where the thin null shell lies at $v^+ = 0$ separating the inside Minkowski from the Schwarzchild exterior, this coordinatization can be computed (by ray tracing) in terms of the Eddington-Finkelstein ingoing coordinates to give $v^+ = v$ and
\begin{multline}\label{va}
v^{-}(v,r)\mathrm{=}
\left
\{
\begin{array}{c}
v-2r \qquad \qquad\qquad\qquad\textrm{for}\ v<0\\
-2r_{s}\Big[1+W\left(\delta\,e^{\delta-\kappa v}\right)\Big]\quad\quad\textrm{for}\ v\geq0,
\end{array}
\right.
\end{multline}
where $\delta\equiv r/r_{s}-1$ and $W(z)$ is the Lambert $W$-function. The metric in these coordinates is given  (with $a \equiv 1 + \kappa\,v^-$) by,
\be
\label{nullmetric}
\hspace{-12pt}\small{ds^2 = - C(v^+,v^-) dv^+dv^- + r^2d\Omega^2;\hspace{8pt}
C(v^+,v^-) =  1+\left(-1 + \f{a-1}{a}\f{W(-a e^{-a + \kappa v^+})}{1+W(-a e^{-a + \kappa v^+})}\right)\Theta(v^+).}
\ee
 
\item[(2)] {\bf Field setup:} The minimally coupled massless scalar field in the $s$-wave approximation at $x$ is given by
\be
\f{\phi(x)}{r}\sim\lim_{r\rightarrow\infty}\f{\phi(v_{+}(x),r)-\phi(v_{-}(x),r)}{r},
\ee
that is, superposition of two spherical waves one impinging directly and one that comes after a reflection from $r=0$. This also gives the interpretation to the coordinates $(v^+,v^-)$ as being tied to the eikonal approximation.     

\item[(3)] {\bf Detector response:} The response of a detector \cite{birrel} moving on a general trajectory $\gamma(\tau)$ can be written as
\be
\dot{\R}(\Omega)=2\,\textrm{Re}\,\int_{0}^{\infty}ds\, e^{-i\Omega s} G\big(\gamma(\tau),\gamma(\tau-s)\big)
\ee
where the Wightmann function, $G(x,y)$ takes the logarithmic form \cite{Helfer03},
\be\label{2d}
G(x,y)\propto\ln\Big(\big(v_{+}(x)-v_{+}(y)-i0\big)\big(v_{-}(x)-v_{-}(y)-i0\big)\Big)
\ee
for $s$-waves. This splits the response into two: $\R = \R_{+} + \R_{-}$ decoupling the ingoing and outgoing modes giving \emph{two temperatures}! The response function in its generality is difficult to compute analytically.
However, if $v(\tau)$ is any solution of the differential equation, 
\be
\ddot{v}(\tau)=-k(\tau)\dot{v}(\tau)
\ee
with $k(\tau)$ as the slowly varying function, then we can consider the adiabatic expansion of the response functional, which gives, 
\begin{align}
\dot{\R}(\tau,\Omega; k = 2\pi T\,] = \f{2\pi}{\Omega(e^{\Omega/T}-1)}\Big(\,1\,+\, 
\f{\dot{T}(\tau)}{T(\tau)^{2}}\f{\Omega}{8\pi T(\tau)}\f{e^{\Omega/2T(\tau)}}{\sinh^{2}(\Omega/2T(\tau))}+\cdots\Big).
\end{align}
That is, all observers, asymptotic or non-asymptotic will perceive a \emph{quasi-thermal} spectrum at 
\be
T_{\pm}=\f{1}{2\pi}\Big|\f{\ddot{v}^{\pm}}{\dot{v}^{\pm}}\Big|
\ee
which is related to the `peeling' condition. It becomes proper thermal temperature in the UV limit ($\Omega\rightarrow\infty$) or under adiabatic condition, $\eta_{\pm}\equiv\Big|\dot{T}_{\pm}/T_{\pm}^{2}\Big|\ll1$.

\item[(4)] {\bf Energy density and Fluxes:} For an alternate prescription, we can compute the RSET of the quantum field in $(1+1)$ dimensions from 
\begin{align}
\langle T_{uu (vv)}\rangle=-{\f{1}{12\pi}}C^{1/2} \pp^2_{u(v)} 
C^{-1/2};\hspace{15pt} \langle T_{uv}\rangle=
{\f{1}{24\pi}}\pp_u\pp_v\ln C.
\end{align}
For $s$-waves, the $(1+3)$ dimensional counterpart is usually considered \cite{Brout95} by multiplying the above expressions with the inverse square distance $(1/4\pi r^2)$. Given the velocity of the observer $u^a$ and a vector $n^a$ such that $n_au^a = 0$, we can construct two local invariant observables \emph{energy density} and \emph{flux} as $U = \langle T_{ab}\rangle u^a u^b$ and $F = \langle T_{ab}\rangle u^a n^b$ respectively and which depend explicitly on the observer's kinematics. These have the advantage over detector response in being local (not dependent on the past trajectory) and exact (no need of adiabatic expansion).
\end{itemize}

\begin{table}[b!]
\begin{center}
\begin{tabular}{@{}ccc@{}}
\br
\multicolumn{1}{l}{} Eff. Temperatures & Asymptotic                        & At horizon crossing \\ \mr
$T_-$                & $T_{H}\Big(E+\sqrt{E^{2}-1}\Big)$ & $4ET_{H}$           \\
$T_+$                & 0                                 &  $T_H/2E$  \\
\br                 
\end{tabular}
\end{center}
\caption{\label{radialtemp}Effective temperatures for a detector on a radially infalling trajectory specified by energy per unit rest mass, $E$ in the asymptotic and near-horizon limits.}
\end{table}

\section{Findings}
\begin{figure}[t!]
\centering
\includegraphics[scale=0.65]{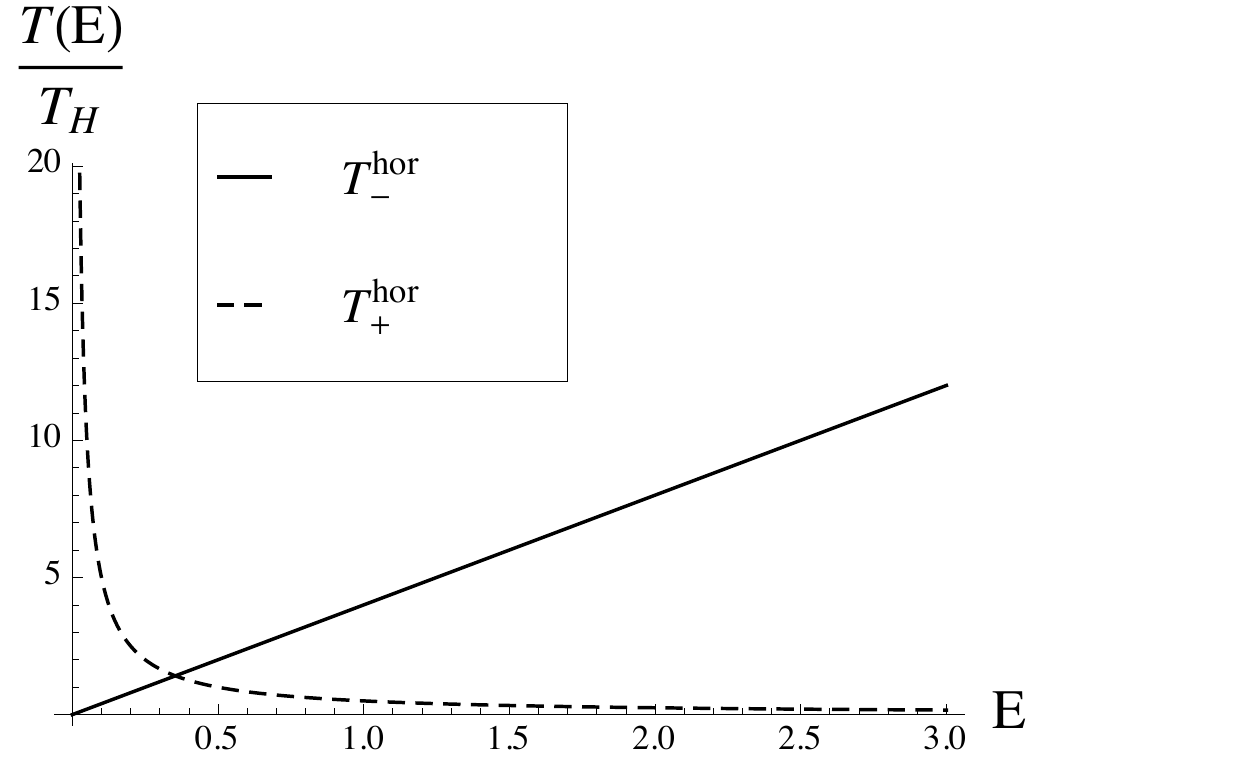}\hspace{3pt}
\includegraphics[scale=0.59]{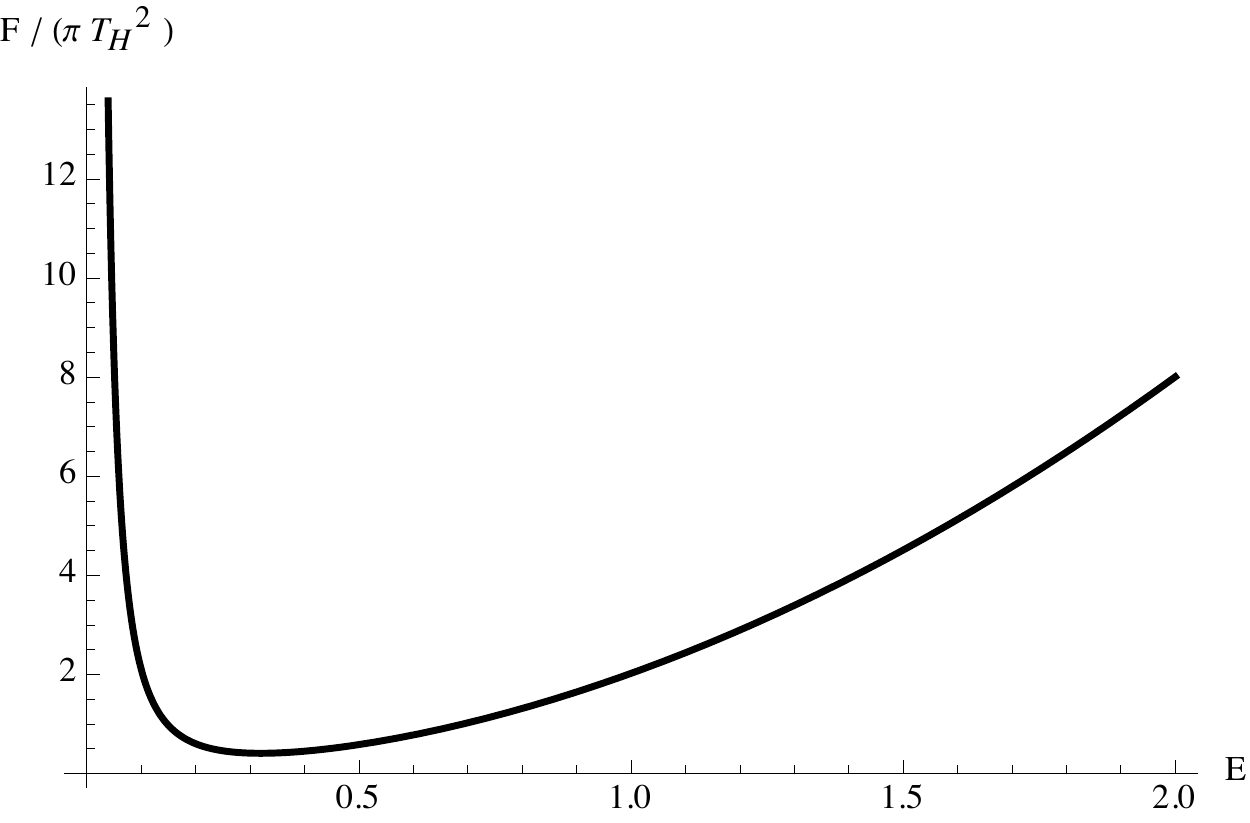}
\caption{{\bf Left.} We have the plot of $T_\pm$ in the near horizon limit which shows dependence on energy and the dominance of ingoing mode ($T_+$) over outgoing ($T_-$) for $E<1/\sqrt{8}$. {\bf Right.} We have the flux in the near horizon region confirming the effective temperature results.}
\label{inoutnearh}
\end{figure}

With the above procedure, the usual Hawking effect for an asymptotic observer at rest relative to the hole can be recovered as follows. For this observer, $v^{-}(\gamma(\tau))$ satisfies
\be
-\f{\ddot{v}^{-}(\gamma(\tau))}{\dot{v}^{-}(\gamma(\tau))}=\kappa
\ee
and subsequently 
\be
v^{-}(\tau)-v^{-}(\tau-s)\propto e^{-\kappa(\tau-s)/2}\sinh(\kappa s/2)
\ee
which recovers \emph{thermal spectrum} at $T_- = T_{H}=\kappa/2\pi$
\be\label{thermalspectrum}
\dot{\R}_{-}(\Omega)\propto {(\Omega(e^{2\pi\Omega/\kappa}-1))}^{-1}.
\ee
while $T_+ = 0$ in this case. For a non-asymptotic static observer at some finite radius $r>r_s$ from the hole, the temperature gets scaled by the redshift factor, $z_r = (1- r_s/r)^{-1/2}$. While many other trajectories can be considered, we shall look at the results for the radially infalling observer characterized by energy per unit rest mass, $E$ or the initial radius from where the infall begins, $r_i$. The analysis of the effective temperatures here then shows a dependence on $E$ and is presented in Table~\ref{radialtemp} (these results are also seen to hold in the case of time-like collapse\cite{ss2014b}). The non-stationarity of the regime shows in having two temperatures now. The temperature $T_-$ presents itself with a doppler shift for $E\neq1$ and the result in the near-horizon limit matches with ref. \cite{Barbado11} for $E=1$, i.e., for the detector dropped in from infinity. The surprise is that in the near-horizon region $T_+$ (incoming modes) dominates over $T_-$ (outgoing modes) for $E\rightarrow0$ (see Fig~\ref{inoutnearh} Left) which implies a duality between the asymptotic and the near-horizon observer in that the near-horizon observer sees a dominant and high radiation from the \emph{sky} \cite{matteo}. However, it is to be noted that the adiabaticity in these cases is $\calO(1)$ and hence the temperatures can only be interpreted in the UV limit.      

\begin{figure}[t!]
\centering
\label{infallUF}
\includegraphics[scale=0.6]{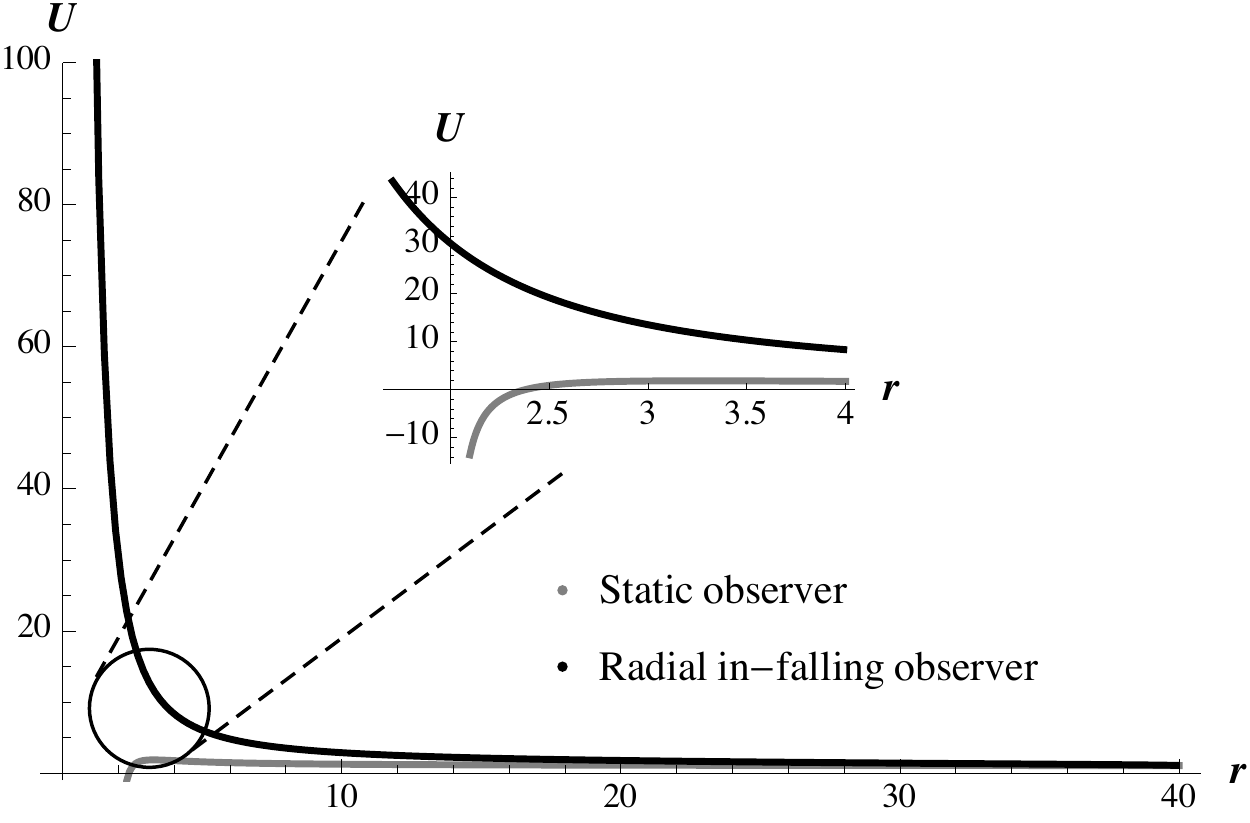}\hspace{15pt}
\includegraphics[scale=0.6]{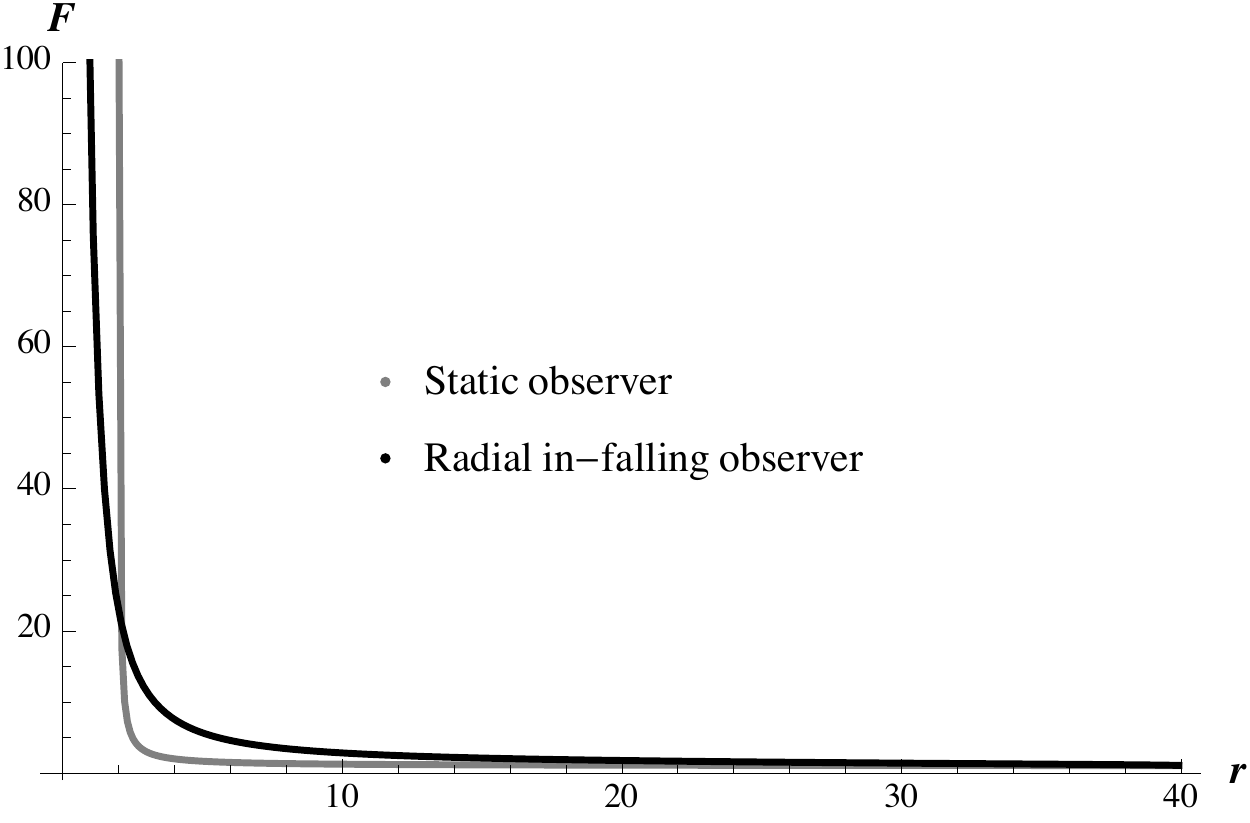}
\caption{The plots show the increase in energy density (left) and flux (right) as the infall occurs in comparison with those perceived by the static observers at the same radii. Both the observables are finite at the horizon crossing (we have taken $r_s = 2$ or $M=1$ here) and diverge only at the eventual singularity. For the static observers, as we go near the horizon, the flux diverges at the horizon due to infinite acceleration. The energy density, however, reaches a maximum and then decrease to become negative and diverges negatively at the horizon.}
\end{figure}

From another angle, we can compute the energy density and fluxes for various observers. For the static observers in the late-time limit, we have,      
\begin{align}
U = \f{\pi T_H^2}{12}\left(1-\f{2\, r_s^4}{r^4}\right)\left(1-\f{r_s}{r}\right)^{-1};\hspace{10pt}
F = \f{\pi T_H^2}{12}\left(1-\f{r_s}{r}\right)^{-1}.
\end{align}
These expressions reduce to the Hawking energy density and flux for the asymptotic observers with $r \rightarrow \infty$ and otherwise the temperature is given by the Tolman blueshift factor. The \emph{onset} of thermality depends on the location of the observer. It is earlier for the near-horizon observers and late for those away from the horizon. The flux diverges at the horizon. The energy density presents a contrast that in the near-horizon region it reaches a maximum positive value, then decreases and becomes negative (Fig. \ref{infallUF}). This corresponds to negative energy fluxes directed into the black hole which have been discussed before (see ref. \cite{Ford93}). 

For the radially infalling observer, the energy density and flux increases (see Fig.~\ref{infallUF}) as the infall occurs and is \emph{finite} and \emph{regular} at the horizon for the infall beginning far from the horizon and diverges only at the eventual singularity. At the horizon, these observables are given by,
\begin{align}
U_H = \pi T_H^2\left(\f{2}{3} - \f{1}{48 E^2} + 2E^2\right); F_H = \pi T_H^2\left(\f{1}{48 E^2} + 2E^2\right)
\end{align} 
which show agreement with the effective temperatures such that for $E\rightarrow0$, that is, if the infall occurs starts close to the horizon, a high flux is perceived at the horizon crossing (see Fig. \ref{inoutnearh} Right). 

{\bf But why the clicking?} We can show by considering artificial black holes which are flat at the horizon, that the effective temperature due to outgoing modes $T_-$ vanishes at the horizon and also the flux contribution corresponding to it. However, the effect due the ingoing modes remains. This shows that curvature has something to do with the detector response and the flux. When the outgoing modes do not feel the curvature at the horizon, we see no effect but the ingoing modes surely encounter some curvature while travelling to the horizon and account for the observed effect. The varying Riemann curvature (as the infall progresses) gives the clicking at the horizon.  

\section{Conclusions}

We see that using a new set of tools and procedure unravels some new and surprising results in the age old problem of quantum fields in gravitational collapse. The response of an UDW detector can be computed using the adiabatic expansion and hence we see not one but \emph{two} temperatures attributed to ingoing and outgoing modes. This formalism can be applied to many different trajectories. In particular, we see for the infalling radial observers, in the near-horizon limit, the temperature due to ingoing modes dominates over outgoing modes for \emph{slowly} moving detectors. A look at the energy density and flux confirms this. So the punchline is that \emph{if you wish to drop into the black hole, start far off}. Also, now we know, at least semi-classically, that there \emph{is} clicking at the horizon and we need to dig in further for more insights. 

\section*{Acknowledgements}

I thank Thanu Padmanabhan and Dawood Kothawala for various discussions, Matteo Smerlak and Sumanta Chakraborty for the collaborations \cite{ms2013} and \cite{ss2014b} respectively. I also thank IUCAA and the organisers of ERE 2014 for the financial support that allowed me to attend the meeting.


\section*{References}

\begin{thebibliography}{9}

\bibitem{Hawking74} S.W. Hawking, \textit{Nature (London)} \textbf{248}, 30 (1974).

\bibitem{Almheiri13} A. Almheiri, D. Marolf, J. Polchinski and J. Sully, \textit{JHEP} 1302, 062 (2013) [arXiv:1207.3123].

\bibitem{Unruh76} W.G. Unruh, \textit{Phys. Rev. D} \textbf{14}, 870 (1976).

\bibitem{Wilburn11} D. Singleton and S. Wilburn, \textit{Phys. Rev. Lett.} \textbf{107}, 081102 (2011).

\bibitem{Barbado11} L.C. Barbado, C. Barcelo and L.J. Garay, \textit{Class. Quant. Grav.} \textbf{28}, 125021 (2011).

\bibitem{DeWitt79}
B.~S. DeWitt, ``{Quantum gravity: the new synthesis},'' in {\em
  General relativity: an Einstein centenary survey}, S.~W. Hawking and
  W.~Israel, eds., pp.~680--745 Cambridge University Press, 1979.

\bibitem {ms2013} M. Smerlak and S. Singh, \textit{Phys. Rev D} {\bf 88}, 104023 (2013) [arXiv:1304.2858].

\bibitem{ss2014a} S. Singh and S. Chakraborty, \textit{Phys. Rev. D} {\bf 90} 024011 (2014) [arXiv:1404.0684].

\bibitem{ss2014b} S. Chakraborty, S. Singh and T. Padmanabhan, work in progress. 

\bibitem{birrel} N.~D.~Birrell and P.~C.~W.~Davies, {\sl Quantum Fields in Curved Space}\/ (Cambridge University Press, 1982).

\bibitem{Helfer03} A. D. Helfer, \textit{Rept. Prog. Phys.} {\bf 66}, 943 (2003). 

\bibitem {Brout95} R.~Brout et al, \textit{Phys. Rept.} {\bf260}, 329 (1995) [arXiv:0710.4345].

\bibitem{matteo} M. Smerlak, Int. J. Mod. Phys. D Vol. 22, No. 12, 1342019 (2013).

\bibitem{Ford93} L. H. Ford and T. A. Roman, \textit{Phys. Rev. D} {\bf 48}, 776 (1993) [arXiv:gr-qc/9303038].


\end{thebibliography}
\end{document}